\documentclass[12pt,a4paper]{article}
\usepackage[left=2.5cm,top=2.5cm,right=2.5cm,bottom=2.5cm]{geometry}
\usepackage[utf8x]{inputenc}
\usepackage{ucs}
\usepackage{authblk}
\usepackage[authoryear]{natbib}
\usepackage{amsmath,amsfonts,bm,amssymb,latexsym,amsthm}
\usepackage{graphicx, graphics}
\usepackage{multirow}
\usepackage[authoryear]{natbib}
\usepackage{microtype}
\usepackage[none]{hyphenat}
\usepackage{setspace} 
\usepackage{array}
\usepackage{amsfonts}
\usepackage[bottom]{footmisc}
\usepackage{color}
\usepackage{url}
\usepackage{float}
\usepackage{rotating}
\usepackage{morefloats}
\usepackage{booktabs}
\usepackage{multirow}
\usepackage[para,online,flushleft]{threeparttable}
\usepackage[justification=centering,font=small]{caption}

\usepackage{hyperref}
\hypersetup{colorlinks, breaklinks, linkcolor=[rgb]{0,0.3,1}, citecolor=[rgb]{0,0.3,1}, urlcolor=[rgb]{0,0.3,1} }

\title{Diversification, economies of scope, and exports growth of Chinese firms\footnote{We thank the Institute for Advanced Research of Shanghai, University of Finance and Economics for providing access to the data. We also thank Giulio Bottazzi and Emanuele Pugliese for providing useful comments and discussions, and the participants of the Workshop ``Crisis, Inequality and Development" 2016 (Shanghai, China), EMAEE 2017 (Strasbourg, France), and MEIDE 2017 (Montevideo, Uruguay).  Le Li gratefully acknowledges the research support by the IBIMET-CNR (grant CrisisLab-ProCoPe).}}

\author{Mercedes Campi\thanks{\textit{Corresponding author}. CONICET - University of Buenos Aires, Faculty of Economics, IIEP-Baires, Buenos Aires, Argentina. mmcampi@gmail.com}, Marco Due\~nas\thanks{Department of Economics, International Trade and Social Policy -- Universidad de Bogot\'a Jorge Tadeo Lozano. marcoa.duenase@utadeo.edu.co}, Le Li\thanks{IBIMET-CNR \& Sant'Anna School of Advanced Studies, Pisa, Italy. lile7k@gmail.com}, and Huabin Wu\thanks{The Institute for Advanced Research of Shanghai University of Finance and Economics, Shanghai, China. wu.huabin@sufe.edu.cn}}

\date{December 22, 2017}

\begin{document}
\maketitle

\begin{abstract}
In the 1990s, China started a process of structural reforms and of trade liberalization, which was followed by the accession to the World Trade Organization (WTO) in 2001. In this paper, we analyze trade patterns of Chinese firms for the period 2000-2006, characterized by a notable increase in exports volumes. Theoretically, in a more open economy, firms are expected to move from the production of a set of less-competitive products towards more internationally competitive ones, which implies specialization. We study several stylized facts on the distribution of Chinese firms trade and growth rates, and we analyze whether firms have diversified or specialized their trade patterns between 2000 and 2006. We show that Chinese export patterns are very heterogeneous, that the volatility of growth rates depends on the level of exports, and that volatility is stronger after trade liberalization. Both, diversification in products and destinations have a positive impact on trade growth, but diversification of destinations has a stronger effect. We conclude that the success of Chinese exports is not only due to an increase in the intensive margin, related to the existence of economies of scale, but also due to an increase in the extensive margin, related to the existence of economies of scope.
\end{abstract}

\noindent \textbf{Keywords:} Industrial dynamics; Margins of trade; Diversification and specialization; Economies of scope

\noindent \textbf{JEL Codes:} F14; F61; L25

\newpage

\section{Introduction}

In the 1990s, China started a process of trade liberalization, along with several reforms across a wide variety of sectors, which was finally followed by the accession to the World Trade Organization (WTO) in 2001. This event is expected to influence the behavior of Chinese exporting firms. 

Recent theories of international trade predict that facing trade liberalization, firms will: (i) reduce the quantity of products that they export, (ii) intensify the volume of exports of a limited number of products, and (iii) increase their market shares on this reduced number of products \citep[see, for example,][]{melitz2003, melitz_helpman, bernard_multi_swit, bernard_multi_lib}. In addition, \cite{pietronero_diversification} argue that while competitiveness at the country level is mainly driven by diversification of productive systems, firms' competitiveness is mainly a matter of specialization. But also, at the country level, the effect of liberalization on trade diversification is likely to depend on the income level of countries. For middle-income countries, some authors find a strong diversification trend after trade liberalization, particularly strong in the five years following liberalization \citep{carrere}.

Along with the increasing interest in how liberalization affects diversification patterns and exports, a broad literature analyses how exports and diversification affect productivity and growth, both at the country level \cite[see, for example,][]{hidalgo, hausmann, pietronero} and at the firm level \cite[see][for a review]{wagner}. 

Firms can diversify on their destinations because this can help them to stabilize exports \citep{kim}. The effect of product diversification is less clear, although a theoretical explanation for the existence of multi-product firms is the reduction of risk that can be reached by diversifiying across product markets, which implies a negative relationship between product diversification and the variability of sales the firm level \citep{lipczynski}. However, \cite{wagner2014} finds that profits tend to be larger in German firms with less diversified export sales over goods and in firms with more diversified export sales over destination countries. 

There is less available evidence on how diversification affects the growth of exports. 
However, if there are economies of scope, diversification on both products and destinations can lead to an increase in the level of sales. In addition, diversification as well as competitiveness depend on technological and organizational capabilities \citep{dosi_pavitt_soete}. For a number of reasons, developing capabilities to export a new product might be more difficult for a given firm than being able to arrive to a new destination with the same product, particularly, if we consider that industries present stickiness in the reallocation of resources from one sector to another. In this sense, we expect diversification in destinations to have a stronger effect on export growth than product diversification. 

In this paper, we investigate the links between diversification over products and destinations and the growth of exports of Chinese exporting firms in a context of trade liberalization and an impressive increase in total exports at a 25.39\% annual growth rate. We use a data set from the Chinese Custom Office that contains information on the volumes and values of exported goods and the destination of exports of Chinese firs for the period 2000-2006. 

Firstly, we analyze the statistical properties of the distribution of Chinese firms exports and their growth rates. Next, we study specialization and diversification patterns of exports in the context of trade liberalization. To do this, we analyze how the intensive and extensive margins of trade explain the evolution of trade during this period and we investigate whether firms exhibit economies of scope both considering products and destinations. Finally, we explore whether exports growth of Chinese exporting firms can be explained by an increase in their traded products and in the total number of destinations of  exports, controlling for several firm characteristics. 

The statistical analysis revealed a high heterogeneity of Chinese trading firms. Many relatively small exporting firms coexist with a few very large exporters. Also, exporting firms are often affected by extreme events (high negative and positive growth rates of trade). Trading firms of different size are similarly affected by negative growth rates, but smaller firms have higher probabilities of growing more than medium and large firms. This implies the existence of a negative relation between size and volatility. Finally, we observe that, after trade liberalization, firms face more frequently stronger negative shocks.

The analysis showed that exports growth at the firm level is due to increases in the intensive and extensive margins of trade, and also to both. Chinese exports have heterogeneous patterns in their diversification dynamics, both in products and destinations. In contrast to the expectations of recent international trade theory, there is no significant evidence of a general process towards specialization after trade liberalization. We observe that the exports of Chinese firms exhibit economies of scope both considering products and destinations. Finally, we observe that diversifying in products and, especially, in destinations increase growth rates of exports, and that more diversified firms in both aspects grow more compared to more specialized firms.  

The rest of the paper is organized as follows. Section~\ref{moti} presents the motivation and the literature review. Section~\ref{data} explains the data and presents a statistical analysis of the distribution of Chinese firms exports and the growth rates of exports. Section~\ref{diver} investigates the role of the intensive and extensive margins of trade and the existence of economies of scope, and presents the econometric estimations of the effect of diversification on trade growth. Finally, Section~\ref{conclu} concludes.


\section{Motivation and literature review}\label{moti}

In the 1990s, China started a dramatic liberalization undertaking several measures, which took place in a period of extraordinary growth in trade and output. While the accession to the WTO in 2001 was the result of this process, it also involved several other reforms across a wide range of sectors in China \citep{ianchovichina2001WTO}. These structural reforms are reflected in several important changes both at the macro and micro level. Between 2000 and 2006, Chinese exports increased at a 25.39\% annual growth rate, private-owned enterprises increased their share in total exports from 0.84\% to 17.56\%, joint ventures and foreign-owned increased their share from 48.45\% to 58.82\%, while state- and collective-owned enterprises decreased their share in total exports from 50.71\% to 23.62\%. 

At the micro level, considering the heterogeneity of firms, we might expect different effects of trade liberalization on firms performance. Several recent models of international trade theory postulate that a more open economy will affect the capability of firms of exporting new products as well as the range of products they specialized on \citep[see, for example][]{melitz2003, melitz_helpman, bernard_multi_swit, bernard_multi_lib}. 

These models predict that trade liberalization will come along with an increasing competition and a decrease in trade costs. Facing an increasing competition, firms are expected to move from the production of a set of less-competitive products towards more internationally competitive ones. Moreover, when trade costs decrease as a consequence of trade liberalization, more productive firms are expected to enter international markets or increase their export shares in their total sales \citep{melitz_helpman}. In brief, theoretical models predict that in the context of trade liberalization, firms will: (i) reduce the quantity of products that they export; (ii) intensify the volume of exports of a limited number of products; and (iii) increase their market share on this reduced number of products. This implies a reduction in the extensive margin of trade and an increase in the intensive margin of trade at the firm level.

In these models, firms' differential efficiency is considered as the main determinant of their participation in the international market \citep{bernard2007}. Instead, an evolutionary perspective departs from the idea that there is persistent heterogeneity among firms and countries, a systematic processes of competitive selection among them, and stickiness in the reallocation of resources from one sector to another \citep{dosi_pavitt_soete}. Thus, international competitiveness responds to differences in variable costs but also to wide differences in technological and organizational capabilities, which shape trade patterns within sectors and countries. \cite{dosi_grazzi} showed that  at the micro level the probability of being an exporter as well as the capacity to increase exports are positively correlated with investments and patents. In addition, several studies indicate that differences in firms performance are highly correlated with their exporting activities \citep{roberts, mathew}. 

At a macro level, there is no consensus on the effect of the WTO on trade. While \cite{rose_2004} found little evidence that countries becoming members or belonging to the General Agreement on Tariffs and Trade (GATT) or the WTO changed their trade patterns compared with those who are not members, \cite{subramanian-wei} found that the WTO has had a positive but uneven impact on trade. 

For China and Chinese firms trade patterns the evidence is limited and mixed. \cite{ianchovichina2001WTO} found that China's major trading partners gained from accession, while some competing countries suffered smaller losses. \cite{rodrik} argues that the success of China's exports owes more to government policies than to comparative advantage and free markets. He claims that, as a result of these policies, China has shifted towards an export basket that is significantly more sophisticated than the one expected for countries at its income level. He also argues that this trade sophistication has been an important determinant of China's rapid growth.

\cite{amiti2008trade} studied the over five-times export growth of China between 1992 and 2005. They found that China's export structure has changed dramatically, with a decline in agriculture and apparel, and growing shares of electronics and machinery. However, when they exclude processing trade, the content of China's manufacturing exports remains practically unchanged. This implies that the seeming shift towards more sophisticated products is not verified once they consider the content of imported inputs that are assembled and then exported. They also found evidence of an increasing specialization along with export growth. Finally, they found that export growth derives mainly from the intensive margin (growth of existing products) rather than from the extensive margin (new products).

Several other authors have highlighted that processing trade is behind the apparent sophistication of Chinese exports \citep[see, for example][]{yao2009, koopman, xing}. However, other authors argue that although sophistication of exports is not deep if processing trade is excluded, there has still being a process of change that resulted in economic growth. \cite{jarreau2012export} studied the effect of export sophistication on economic performance of different regions within China between 1997 and 2009. They found substantial variation in export sophistication at the province and prefecture level. They showed that regions specializing in more sophisticated goods subsequently grow faster. But, their results suggest that gains are limited to the ordinary export activities undertaken by domestic firms, given that no direct gains result from either processing trade activities or foreign firms.

Also, other authors suggest that not all the sophistication of Chinese exports is due to an increase in processing trade. \cite{wang} found that there are relevant regional variations in the use of processing trade. They argue that there exist cross-city differences in human capital, which are linked to cross-city differences in the sophistication of the export structure. They also argue that the increasing sophistication owes to the government promotions through high-tech and economic development zones, and that foreign investment has been conducive to greater product sophistication in China.

\cite{manova2009china} analyzed Chinese trade flows at the firm level for the years 2003 to 2005. They confirmed several stylized facts that also characterize firms trade in other countries. They showed that the bulk of exports and imports are captured by a few multi-product firms that transact with a large number of countries. They also observe that, compared to private domestic firms, foreign affiliates, and sino-foreign joint ventures import more products from more countries, but export fewer products to fewer destinations. Moreover, they found that the relationship between the intensive and extensive margins of trade is non-monotonic, differs between exporters and importers, and depends on the ownership type of the firm. They also found that firms frequently exit and re-enter into trade and regularly change their product mix and trade partners, but foreign firms exhibit less churning. Finally, they showed that the growth in Chinese exports between 2003 and 2005 was mainly driven by deepening and broadening of trade relationships by surviving firms.


While the composition of trade and diversification patters have been analyzed by a large number of studies, to our knowledge, there are no analysis of how specialization or diversification of Chinese exporting firms affect the growth of exports. Recent trade theories have focused on the exploration of the linkages between productivity and trade, highlighting complex relationships between trade diversification and productivity \cite[see][for a review]{carrere}. But, diversification or specialization can also have an effect on the growth of trade, particularly if there exist economies of scale or economies of scope. 

\section{Data and statistical analysis}\label{data}

Our database contains exports and imports of the universe of Chinese firms for the period 2000 to 2006, which are collected from trade records in the Chinese Customs Office. 
The completeness of the data is confirmed with a simple exercise. According to the National Statistics Office, total exports of China in 2001 was 266.10 and total imports 243.55 billion dollars.\footnote{See \url{http://www.customs.gov.cn/publish/portal0/tab44604/module109000/info1177.htm} (in Chinese), accessed on December 2017.} The aggregation of our firm level data provides a value of 266.66 billion dollars for total exports and 243.57 billion dollars for total imports. We have used deflators from the US Bureau of Labor Statistics to have the data in constant US dollars.\footnote{See \url{http://www.bls.gov/news.release/ximpim.t06.htm}, accessed on April 2016.}

The data set includes information on the volume of exported and imported products and the value in US dollars, the date of the shipment, the HS code at 8 digits of disaggregation, the custom that reports the shipment (there are 41 customs across China), a firm identifier code, the city of the firm, type of shipment (general trade, processing trade, etc.), country of destination for exports and country of origin for imports, and the ownership type of the firms.

In order to use these data, we first took the following data cleaning procedures. First of all, we excluded records with missing data. 
Secondly, we only kept observations for ordinary trade and processing trade (adding 4 different types of processing and assembling trade).\footnote{Although there are 19 types of trade, these two broad categories include between 96 and 98\% of total trade each year. The categories that we excluded are not useful for our purposes because they include special deals such as compensation trade, duty free foreign exports, or donated materials.} Ordinary trade refers to the export of a product mainly produced with Chinese inputs. Instead, processing trade includes products with a high content of imported inputs (raw materials, parts and components, accessories, and packaging materials) that are processed or assembled by firms and then exported. For our period of study, processing trade is more than half of total Chinese exports (57.81\% in 2000 and 54.50\% in 2006).


Finally, the data have 8 different ownership types.
We grouped these types into three main categories:(i) stated-owned and collective-owned enterprises, (ii) private-owned enterprises (which include private enterprises, and individual industrial and commercial households), and (iii) foreign-owned enterprises (including sino-foreign contractual joint ventures and sino-foreign equity joint ventures). The shares of these three groups in total exports has changed between 2000 and 2006, increasing the shares of private-owned and foreign-owned enterprises and decreasing the share of state-owned enterprises. 

We define $X_{i,t}$ as total exports of firms, where $X$ stands for the total value of exports for firm $i$ in year $t$. The growth of firms exports ($r$) is the log difference of total exports in the consecutive year: 
\begin{equation}
 \label{eq:growth_x}
 g_{i,t}=\ln{(X_{i,t})}-\ln{(X_{i,t-1})}.
\end{equation}

We define the number of exported products using a 6-digit code within the Harmonized System classification. We take the number of different products exported and the number of foreign destinations as a proxy for firm scope as in \cite{bee}.

Table~\ref{tab:summary} reports the shares of firms and exports for the years 2000 and 2006 for the full sample and for different sub-samples based on the number of exported products and destinations of firms. 

\begin{table}[h!]
  \begin{center}
  \renewcommand{\arraystretch}{1.2}
  \begin{footnotesize}
  \caption{Summary statistics: share (in \%) of the number of firms and total exports (in millions) for the full sample and different sub-samples for 2000 and 2006}\label{tab:summary}
\begin{tabular}{l c c c c }
\toprule
 & \multicolumn{2}{c}{Number of firms} & \multicolumn{2}{c}{Total Exports}\\
  \cmidrule(lr){2-3}
   \cmidrule(lr){4-5}
 & 2000 & 2006 & 2000 & 2006  \\
 \midrule
\multicolumn{5}{l}{\textit{Shares by product diversification}} \\
\hline
Single product  & 24.22 & 24.20 & 6.93 & 6.05 \\  
Multi-product  & 75.78 & 75.80 & 93.07 & 93.95 \\ 
 \midrule
\multicolumn{5}{l}{\textit{Shares by destination diversification}} \\
\hline
Single destination  & 31.54 & 27.37 & 7.65 & 4.45 \\
Multiple destination  & 68.46 & 72.63 & 92.35 & 95.55 \\ 
 \midrule
\multicolumn{5}{l}{\textit{Shares by joint diversification types}} \\ 
\hline
Multiple product-destination  & 58.05 & 62.51 & 87.42 & 90.97 \\  
Single product-destination  & 13.82 & 14.08 & 2.01 & 1.47 \\  
Multi-product and single destination  & 17.73 & 13.29 & 5.64 & 2.98 \\
Single product and multiple destination  & 10.40 & 10.12 & 4.92 & 4.58 \\
 \midrule
Total  & 46,279 & 166,930 & 225,769 & 865,038 \\ 
\bottomrule
\end{tabular}
\end{footnotesize}
\end{center}
\end{table}

The full sample of firms included 46,279 firms in 2000 and 166,930 in 2006, with 27,415 surviving firms, which indicates a process of entry and exit of exporting firms. We do not observe changes in the shares of single and multi-products firms. In both years, almost a quarter of firms export only one product and around 6-7\% of total exports while 75\% of firms export multiple products and around 93-94\% of total exports. Conversely, the share of firms that export to multiple destinations increased from 68.46\% in 2000 to 72.63\% in 2006, while firms exporting to only one destination decreased from 31.54\% to 27.37\% of total firms. Likewise, if we classified firms according to both diversification in products and destinations, we observe that highly-diversified firms that export more than one product to multiple destinations increased from 58.05\% to 62.51\%, while firms specialized in one product and one destination remained unchanged (14\%) exporting a very low share of total exports (1-2\%). Finally, single-product firms shipping to multiple destinations are around 10\% in both years, while single destination firms exporting multiple products decreased from 17.73\% in 2000 to 13.29\% in 2006. Interestingly, highly-diversified firms accounted for 87.42\% of total sales in 2000 and 90.97\% in 2006.  

A number of studies have documented the presence of regularities in empirical data of trade flows and bilateral trade flows \cite[see, for example][]{melitz_helpman, easterly, fagiolo, big_hits}. The main stylized facts of trade and bilateral trade flows are: (i) sparsity, which implies the presence of many zeros, (ii) skewed distribution at the extensive margin of trade (power law distribution), (iii) concentration at the intensive margin (log-normal distribution), and (iv) high volatility in the growth rates (fat tailed distributions).

Some of these statistical regularities have been also confirmed at the firm level \cite[see, for example][]{bee}. 
Bellow, we carry on a statistical analysis in order to explore whether our firm level data present these empirical regularities. Our time period allows us to study trade patterns in the context of liberalization, so we present the analysis for 2000 and 2006.

\begin{figure}[!ht]
\centering%
\includegraphics[width=0.6\textwidth]{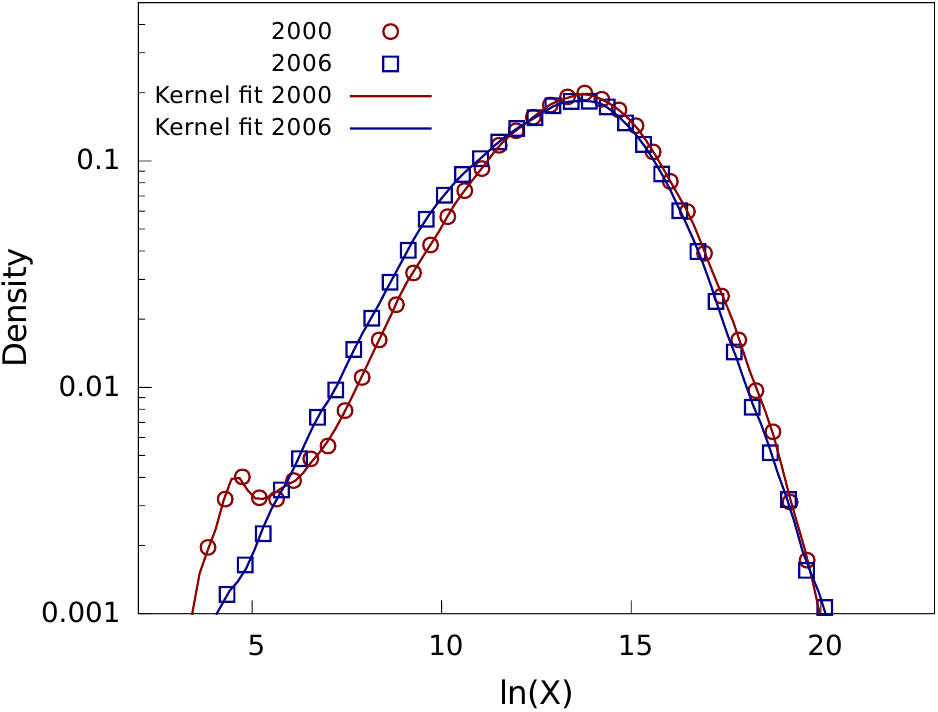}
\caption{Distribution of firms' total exports}
\label{fig:distribution}
\end{figure}

Figure~\ref{fig:distribution} shows the distribution of firms exports for the years 2000 and 2006. Firstly, we observe for both years a wide variability in the size of firms measured by total exports. Both distributions resemble a log-normal, which seem somehow left-skewed but still quite symmetric. These distributions evidence the broad heterogeneity of Chinese firms, with a large number of relatively small exporters that coexist with a few very large exporters.

The empirical literature on trade has shown that firms and also countries grow and decline driven by extreme episodes of expansion and contraction that are relatively frequent \citep{fagiolo}. This is reflected in fat tailed distributions of growth rates across countries and firms, different levels of sectoral disaggregation, and for other measures of size \cite[see, for example,][]{lee_stanley, bottazzi, bottazzi_secchi}. This fat tailed ubiquity in the distribution of growth stresses the existence of strong interdependence or correlating mechanisms that, ultimately, determine growth patterns.

\begin{figure}[!ht]
\centering%
\includegraphics[width=0.6\textwidth]{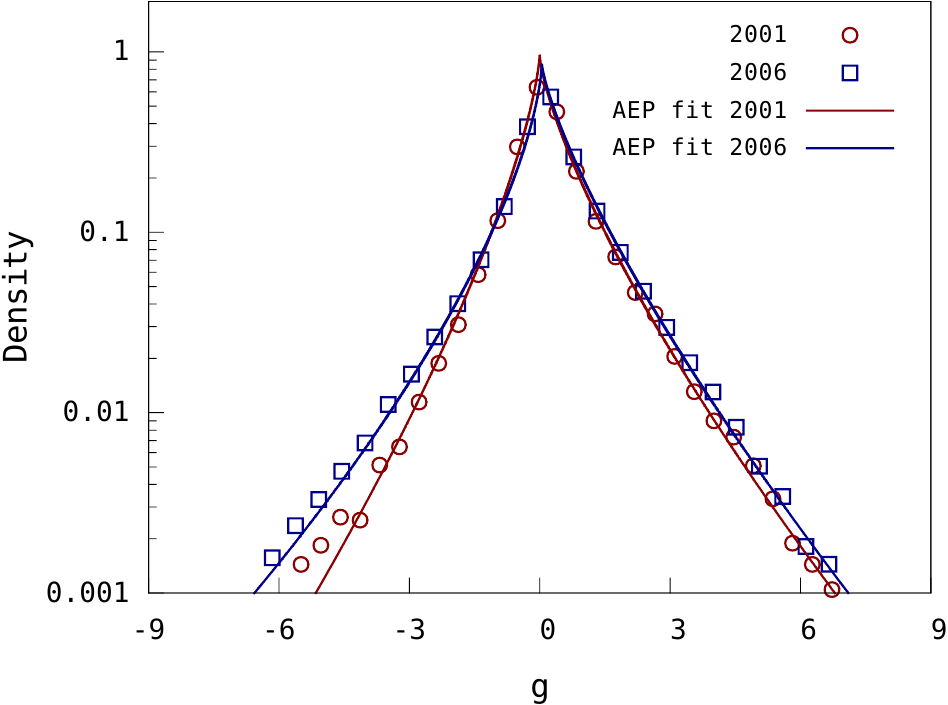}
\caption{Distribution of exports growth of Chinese firms for 2001 and 2006. AEP: asymmetric exponential power}
\label{fig:growth}
\end{figure}

Figure~\ref{fig:growth} shows the distributions of the growth rates of exports for the years 2000-2001 and 2005-2006. We observe fat tails in both sides of the distributions (excess of kurtosis), which implies that the probability of finding extreme events affecting both negatively and positively growth rates is high, compared to a normal distribution. Not surprisingly, the distributions of growth rates are more asymmetric than the distributions of total exports (Figure~\ref{fig:distribution}).

It is interesting to note that export dynamics is quite susceptible to extreme events and present high lumpiness at different levels of aggregation. In fact, firms appear and disappear from one year to the other as we observed in Table~\ref{tab:summary}. Internal and external competition, market diversification, either in products or in destinations, and the constitution or closure of plants are all phenomena that can lead to extreme events.

Interestingly, we observe that in 2006, after trade liberalization, there appears to be more dispersion in the left tail of the distribution.\footnote{For the sake of simplicity, we do not present the estimations for every year, but the distributions present more volatility for all the years compared to the distribution of growth rates for 2000-2001, this is before China became a member of the WTO.} This implies that firms were more frequently affected by negative shocks after 2000, and also, as the left tail is wider, that negative growth rates were higher. 

\begin{figure}[!ht]
\centering%
\includegraphics[width=0.6\textwidth]{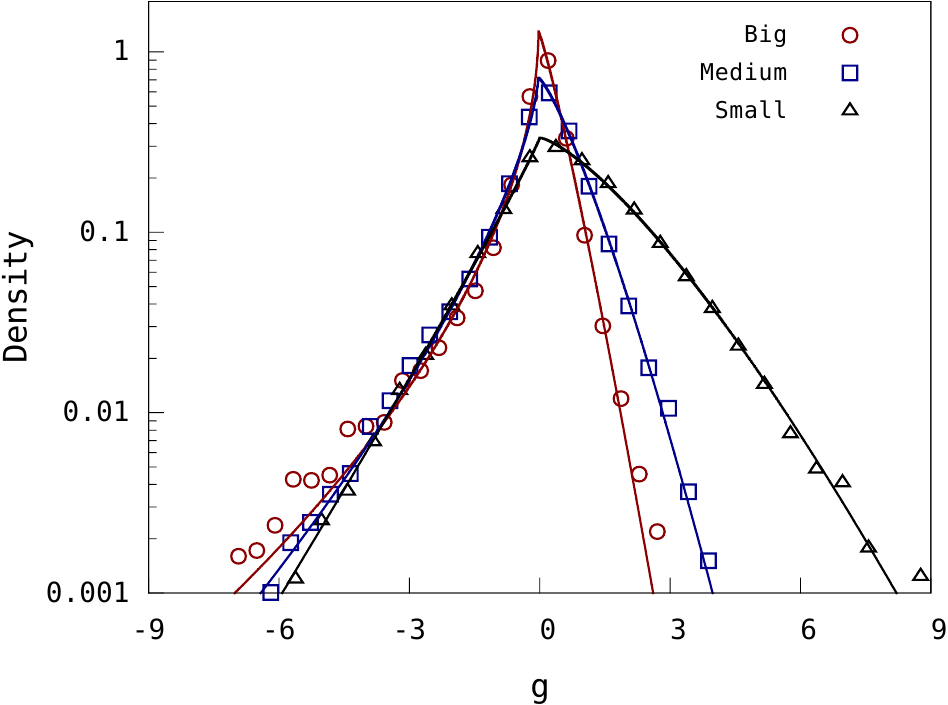}
\caption{Kernel estimation of the empirical firm growth rate distributions for the year 2006 by size binned in three equipopulated categories: big, medium, and small}
\label{fig:growth_binned}
\end{figure}

Figure~\ref{fig:growth_binned} shows the distributions of growth rates for 2006 and for three different equipopulated size bins of exporting firms: small, medium and large. We observe in the right side (positive growth), that small firms (that export little) grow more than medium and, especially, than large firms. In the left side (negative growth), the probability of facing negative shocks is more uniform for different size bins. Therefore, firms of different size are similarly affected by negative growth rates, but smaller firms have higher probabilities of increasing their exports at high rates compared to medium and large firms. It is interesting to note that the distance between the left and right tails increases with size, which implies that large firms have less volatility in their trade growth rates.


Regaring the ubiquity in the fat tailed distribution of growth rates of firms, \cite{dosi2007} claims that the variance-scale relation depends on the relation between diversification and size. The growth of firms can be explained by both the increase in the existing lines of business, as well as through the diversification of the existing ones. This is also interesting for the growth of trade and refers to the effect of the intensive and extensive margins of trade.

\section{Trade growth: the margins of trade and \\diversification in products and destinations}\label{diver}


In this section, we investigate how product and trade partners diversification have affected Chinese exporting firms. We claim that diversification in products and destinations can positively affect firms trade growth. Firstly, we analyze the intensive and extensive margins of trade. Next, we investigate whether exporting firms present economies of scope related to both product and destination diversification. Finally, we study how diversification in products and destinations affect the growth of firms' exports. 

\subsection{The margins of trade}

In the first place, we analyze the extensive and intensive margins of exports in both destinations and products, which can provide a broad picture of the diversification patterns. For the case of diversification in destinations, the extensive margin for a firm $i$ is defined as the number of exporting destinations $N_{d,i}$, and the intensive margin as the average total exports by destination $X_{i}/N_{d,i}$. In a similar way, in the case of diversification in products, the extensive margin is defined as the number of exporting products $N_{p,i}$, and the intensive margin is defined as the average trade by products $X_{i}/N_{p,i}$. 

We are interested in determining whether firms have specialized or diversified their exports between 2000 and 2006. The question we want to address is whether the evolution of exports and the creation/destruction of markets (products and destinations) have been governed by the intensive margin, the extensive margin, or both. As mentioned, from a theoretical point of view, a consequence of competing in international markets is that firms are expected to specialize in more productive products \citep{melitz_helpman}. Therefore, a possible outcome of trade liberalization is that firms will intensify their exports in some products and destinations. This means that, in the context of increasing competition, the extensive margin will be reduced while the intensive margin will be expanded. 

We select the sample of surviving firms from 2000 to 2006, and for each of them we calculate the ratio of the extensive margin and the ratio of the intensive margin for 2006 over 2000. Correspondingly, when these ratios are greater than one, the extensive and intensive margins of a firm have enlarged. Conversely, if the ratio is less than one, the extensive and intensive margins have shrunk. We do this for both diversification in products and in destinations of exports. Table~\ref{tab:ext_int} shows the percentage of firms with changes in their intensive and extensive margins between 2000 and 2006.

\begin{table}[h!]
  \begin{center}
  \renewcommand{\arraystretch}{1.1}
  \begin{footnotesize}
  \caption{Changes in the intensive margin (IM) and the extensive margin (EM) of exports between 2000 and 2006 for surviving firms in percentage}\label{tab:ext_int}
\begin{tabular}{l c c  }
\toprule
 &Product (\%) & Destination (\%) \\
 \hline
IM enlargement     & 59.14 & 55.25\\ 
EM enlargement    & 47.81 & 54.76\\ 

\hline
\multicolumn{1}{l}{\textit{Joint changes of margins}}\\     
\hline 
EM and IM shrink    & 20.23 & 21.31\\ 
EM shrinks and IM enlarges    & 31.96 & 23.94\\ 
EM enlarges and IM shrinks    & 20.63 & 23.44\\ 
EM and IM enlarge   & 27.17 & 31.31\\
     
\bottomrule
\multicolumn{3}{p{9.7cm}}{\textit{Notes:} The number of firms was 46,279 in 2000 and 166,930 in 2006. We used 27,415 surviving firms to estimate the margins of trade.}
\end{tabular}
\end{footnotesize}
\end{center}
\end{table}

The first thing to notice is the high increase in the number of firms and the relatively small number of surviving firms. In 2000, there were 46,279 exporting firms , which reached, in 2006, 166,930 exporting firms, increasing 3.61 times. The number of surviving firms is 27,415, 59.24\% of the existing firms in 2000 and 16.42\% of total firms in 2006. It is worth mentioning that not necessarily the missing amount of firms actually died, they could have merged with other international affiliates, changed their business name, or also they could have not traded in the international markets but only in the domestic market. However, this reflects a dynamic market in which there is exit and entry of exporting firms. 

Analyzing diversification in products, we observe that 59.14\% of firms were able to increase their intensive margins of exports. In the case of destinations, the conclusion is quite similar, we observe that 55.25\% of firms increase their intensive margins. To some extent, this supports the hypothesis that trade liberalization leads to specialization. However, we also observe a high percentage of firms that diversify their exports through the enlargement of their extensive margin both in products (47.81\%) and destinations (54.76\%). Thus, there is no conclusive evidence of a general process towards specialization.

Also, we analyze the joint changes in the margins of exports. If we consider products, a firm can: (i) shrink both the extensive and intensive margins, meaning that it specializes in a lower number of products and exports less per product (20.23\% of firms), (ii) shrink the extensive margin while enlarging the intensive margin, therefore, specializing in a lower number of products and exporting more per product (31.96\%), this implies specialization --in line with the theoretical expected outcome we discussed above--, (iii) enlarge the extensive margin while shrinking the intensive margin, meaning that it diversifies the exported products but exports less per product (20.63\%), and (iv) enlarge both margins of trade, which means that the firm diversifies its exported products and increases the quantity of exports per product (27.17\%). Also, in the case of destinations, we observe that a relatively similar percentage of firms appear in each of the three first possibilities (21.31, 23.94, and 23.44\%, respectively), but a slightly higher percentage of firms (31.31\%) diversify their destinations and increase the volume of exports to those destinations, enlarging both the intensive and extensive margins.

Notice that, an increase in the intensive margin can indicate the existence of economies of scale. While there is a percentage of firms that specialize and intensify their exports, as theoretically expected in the context of liberalization, the proportion of firms that manage to diversify their exports is not negligible at all. The expansion of the extensive margin derived from an increase in the number of products and destinations for an important part of the population of Chinese exporters might indicate the existence of economies of scope.

Overall, the analysis of the changes on the margins of trade between 2000 and 2006 evidences heterogeneous patterns in the dynamics of exporting firms, both in products and destinations, as expected from an evolutionary perspective. However, it is important to highlight that these heterogeneous changes could be also related with changes made by China in its industrial and foreign trade policies, such as attracting more foreign capital and integrating to global value chains.


\subsection{Economies of scope and diversification in products and \\destinations}

We are interested in studying how diversification affects the growth of exports. Thus, we test whether exports of Chinese firms exhibit economies of scope. In the context of our study, and given the limitations of the data, the economies of scope mean that the number of existing markets of a firm (exported products or destinations of exports) can be determined by the total volume of exports of the firm itself. In addition, the existence of economies of scope imply that the average total cost of production decreases as a result of increasing the number of different products exported or of increasing the number of destinations. 
If Chinese exports have economies of scope, we expect firms to increase their number of products and destinations as their exports grow.

The diversification behavior of firms can be modelled by a birth process, this is by describing how the probability of having a given number of markets changes as the firm grows. \cite{bottazzi2006} showed that the Yule process provides a good characterization of the diversification patterns of the pharmaceutical industry. As presented in \cite{feller}, in this process it is considered that a product can give birth to a new product but cannot die. Therefore, given a small expansion of exports of length $h$ for a firm, each of its markets (products or destinations) has probability $\lambda h + o(h)$ of creating a new market, where the parameter $\lambda$ determines the rate of increase of the population. Assuming that at size $s$ there is no interaction between markets (no instantaneous branching), then for a firm with $n$-markets the probability of diversifying in one market between $s$ and $s+h$ is $n \lambda h + o(h)$. The probability $P_n(s)$ that the number of markets is exactly $n$ satisfies the following differential equation: 
\begin{equation}
 \begin{split}
 \dot{P}_n(s) & =-n\lambda P_n(s) + (n-1)P_{n-1}(s) \text{ with: } n \geq n_0 \\ 
 \dot{P}_n(s) & = -n_0\lambda P_{n_0}(s); 
 \end{split}
\end{equation}
where $n_0>0$ denotes the initial number of markets at size $s_0$. It can be verified that the solution of this equation for $n \geq n_0$ is 
\begin{equation}
P_n(s) = \binom{n-1}{n-n_0} e^{-n_0\lambda (s-s_0)}(1-e^{-\lambda (s-s_0)})^{n-n_0}.
\end{equation}
This is called Yule distribution and the average number of markets (products and destinations) is:
\begin{equation}
\eta(s) = n_0 e^{\lambda (s-s_0)}.
\end{equation}
Given that the parameter $\lambda$ is strictly positive, the number of markets is expected to grow exponentially with firm size. 

\begin{figure}[!ht]
\centering
 \includegraphics[width=0.496\textwidth]{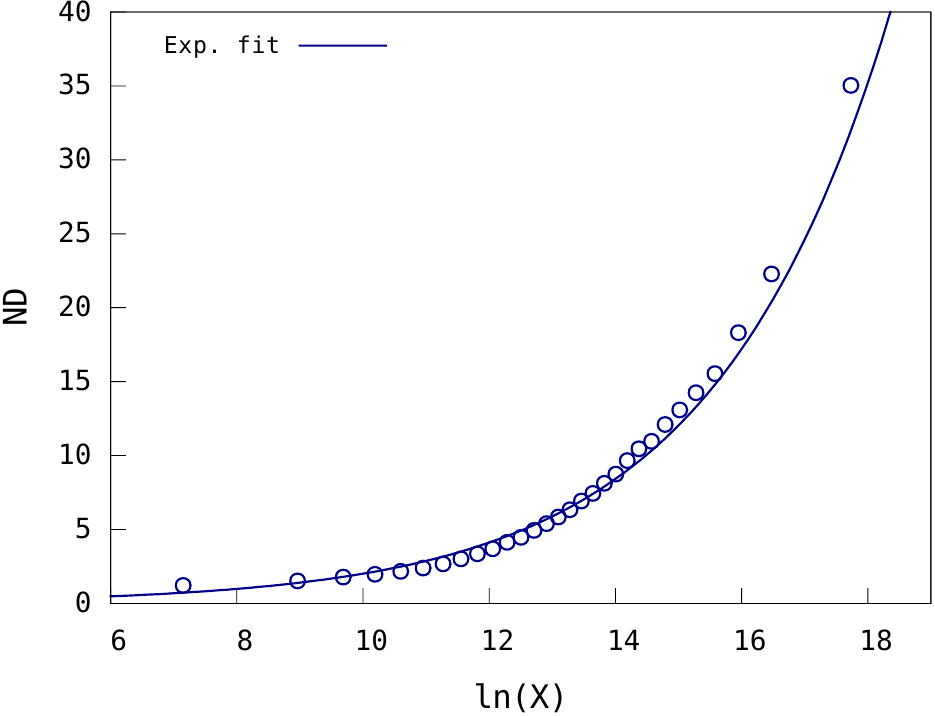}
 \includegraphics[width=0.496\textwidth]{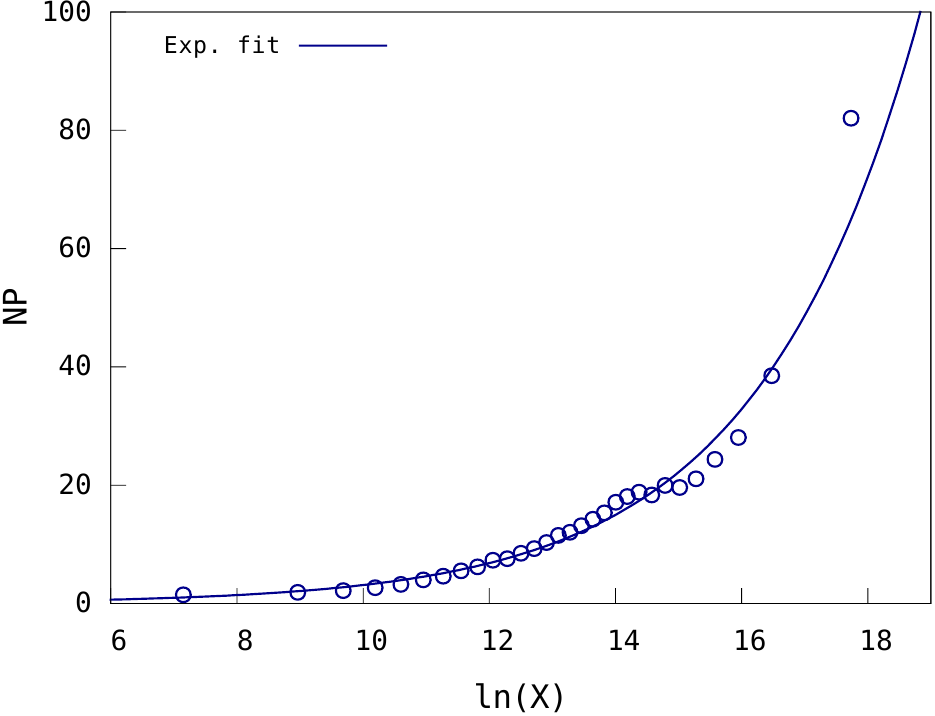}
 \caption{Scatter plots of binned statistics: total exports versus number of destinations (ND) and versus number of products (NP), and exponential fit for 2006}
\label{fig:scope}
\end {figure}

Figure~\ref{fig:scope} (left) shows the average number of destinations of the firms belonging to different size bins against the bin average ($\ln$) total exports. Similarly, Figure~\ref{fig:scope} (right) shows the average number of products of firms belonging to different size bins against the bin average ($\ln$) total exports.\footnote{We present the estimations for the year 2006, but similar relations are estimated for all years. Results are available upon request.} The evidence suggests the above mentioned exponential positive relation between the number of markets and total exports. The continuous line in each figure corresponds to the linear fit between the firm ($\ln$) total exports and the $\ln$ of diversification, accordingly, for destinations and number of products:
\begin{align}
\ln(ND) & \sim \lambda_d \ln(X) + \theta_d, \\
\ln(NP) & \sim \lambda_p \ln(X) + \theta_p. 
\end{align}

The estimated values are: $\lambda_d=0.36(0.01)$ and $\lambda_p=0.39(0.01)$ and $\theta_d=-2.88(0.08)$ and $\theta_p=-2.88(0.09)$. We observe a clear positive exponential relation between the number of destinations/products and the size of the firm, measured in terms of the volume of exports. The existence of economies of scope indicates that diversification of products and destinations can be also positively correlated with firms growth rates and that the effect of creating a new market has diminishing returns as the firm gets bigger.



\subsection{Diversification and the growth of exports}

In order to further investigate the effect of diversification of products and destinations of exports on firms trade growth, we carry out an econometric exercise. Given the evidence of economies of scope on the exports of Chinese firms, we expect diversification to have a positive effect on exports growth rates. Given that diversification depends on the development of capabilities, diversification of destinations might be relatively easier to achieve for a firm compared to diversification of products, which implies the development of a new exportable product. In addition, we include in the estimations several control variables that aim to capture structural changes experienced between 2000 and 2006.

We estimate the following model:
\begin{equation}
\begin{split}
g_{i,t} = \text{ }
          &   \alpha + \phi \cdot g_{i,t-1} + \beta \cdot \ln(X_{i,t-1})+ \theta_{p} \cdot \Delta\ln(NP_{i,t}) + \theta_{d} \cdot \Delta\ln(ND_{i,t}) + \\
          & \omega_u \cdot FirmType_{i,t} + \rho \cdot ProcTrade_{i,t} + \nu_v \cdot Ownership_{i,t} + \\
          & \gamma_v \cdot Ownership_{i,t}\cdot ProcTrade_{i,t} + \tau_t + \epsilon_{i,t}; 
\end{split}
\end{equation}
where the dependent variable $g$ is the growth rates of exports as defined in equation~\eqref{eq:growth_x}, and $g_{i,t-1}$ is the growth rates of exports in $t-1$ (autoregressive term), $\ln(X_{i,t-1})$ is the volume of exports in $t-1$ (an indicator of the catching-up capacity), $\Delta \ln(NP_{i,t})$ is the change in the number of products a firm exports from period $t$ to $t-1$, $\Delta \ln(ND_{i,t})$ is the change in the number of destinations of exports from $t-1$ to $t$, $\omega_u$ controls for a set of four firm diversification types, where $\omega_1$ is a firm that exports multiple product to multiple destinations, this is, highly-diversified firms, $\omega_2$ is a highly-specialized firm, this is, single product and destination firms, $\omega_3$ is a multi-product and single destination firm, and $\omega_4$ is a single product and multiple destination firm, $\rho$ controls if the firm $i$ does processing trade at time $t$, $\nu_v$ controls for a set of three ownership types: with $\nu_1$ includes state-owned and collective-owned enterprises, $\nu_2$ private-owned enterprises, and $\nu_3$ are foreign-owned enterprises, including joint ventures, $\gamma_v$ controls for the interaction between ownership type and processing trade, $\tau_t$ is a set of time-dummies. Finally, $\epsilon_{i,t}$ is the residual term.



We estimate the model with ordinary least squares (OLS) and an asymmetric least absolute deviation (ALAD) estimation methods. The error term is assumed to be normally distributed in the OLS estimations and Laplacian distributed in the ALAD estimations. The ALAD is preferred to OLS when there are outliers and when the distribution of the residuals is non-normal, asymmetric, and has high kurtosis. The assumptions of the ALAD better agree with what we observed in Figure~\ref{fig:growth}. Table~\ref{tab:estimations} shows the results of the estimations of the growth of exports. Models (1-4) show the results of the OLS estimations and models (5-8) report the results of the ALAD estimations.

\begin{table}[t!]
\begin{center}
\caption{The effect of product and trade partner diversification on the growth of exports. \\Econometric estimations}
\resizebox{\textwidth}{!}{
\renewcommand{\arraystretch}{1.3}
\begin{tabular}{lcccccccc} 
\toprule
Estimation Method & \multicolumn{4}{c}{OLS} & \multicolumn{4}{c}{ALAD} \\
 \cmidrule(lr){2-5}
  \cmidrule(lr){6-9}
 Model &	(1)	&	(2)	&	(3)	&	(4) & (5) & (6) & (7) & (8) \\
\hline
Growth$_{(t-1)}$	&	-0.026***	&	-0.026***	&	-0.028***	&	-0.025***	&	0.012***	&	0.020***	&	0.017***	&	0.019***	\\
	&	(0.002)	&	(0.002)	&	(0.002)	&	(0.002)	&	(0.003)	&	(0.001)	&	(0.001)	&	(0.001)	\\
Exp.$_{(t-1)}$	&	-0.063***	&	-0.100***	&	-0.102***	&	-0.103***	&	-0.049***	&	-0.057***	&	-0.060***	&	-0.060***	\\
	&	(0.001)	&	(0.001)	&	(0.001)	&	(0.001)	&	(0.002)	&	(0.001)	&	(0.001)	&	(0.001)	\\
$\Delta$ $ln $ products$_{t}$  	&	0.516***	&	0.476***	&	0.478***	&	0.477***	&	0.374***	&	0.346***	&	0.347***	&	0.347***	\\
	&	(0.004)	&	(0.004)	&	(0.004)	&	(0.004)	&	(0.005)	&	(0.002)	&	(0.002)	&	(0.002)	\\
$\Delta$ $\ln$ trade partners$_{t}$	&	0.698***	&	0.630***	&	0.634***	&	0.632***	&	0.472***	&	0.490***	&	0.491***	&	0.490***	\\
	&	(0.005)	&	(0.004)	&	(0.004)	&	(0.004)	&	(0.006)	&	(0.002)	&	(0.002)	&	(0.002)	\\
Simgle prod.-dest.	&		&	-0.577***	&	-0.564***	&	-0.571***	&		&	-0.277***	&	-0.266***	&	-0.273***	\\
	&		&	(0.009)	&	(0.009)	&	(0.009)	&		&	(0.004)	&	(0.004)	&	(0.004)	\\
Multi-prod. \& single dest. 	&		&	-0.037***	&	-0.026***	&	-0.030***	&		&	-0.008	&	0,000	&	-0.003	\\
	&		&	(0.006)	&	(0.006)	&	(0.006)	&		&	(0.004)	&	(0.004)	&	(0.004)	\\
Single prod. \& multiple dest. 	&		&	-0.135***	&	-0.120***	&	-0.124***	&		&	-0.068***	&	-0.056***	&	-0.059***	\\
	&		&	(0.005)	&	(0.006)	&	(0.005)	&		&	(0.003)	&	(0.003)	&	(0.003)	\\
Processing trade (PT)	&		&	0.123***	&	0.154***	&		&		&	0.043***	&	0.067***	&		\\
	&		&	(0.003)	&	(0.004)	&		&		&	(0.002)	&	(0.002)	&		\\
Private-owned	&		&		&	0.037***	&		&		&		&	0.029***	&		\\
	&		&		&	(0.006)	&		&		&		&	(0.004)	&		\\
Foreign-owned	&		&		&	-0.061***	&		&		&		&	-0.044***	&		\\
	&		&		&	(0.005)	&		&		&		&	(0.003)	&		\\
PT$\times$private-owned	&		&		&		&	0.211***	&		&		&		&	0.116***	\\
	&		&		&		&	(0.008)	&		&		&		&	(0.005)	\\
PT$\times$foreign-owned	&		&		&		&	0.100***	&		&		&		&	0.121***	\\
	&		&		&		&	(0.004)	&		&		&		&	(0.006)	\\
PT$\times$state-owned	&		&		&		&	0.252***	&		&		&		&	0.027***	\\
	&		&		&		&	(0.007)	&		&		&		&	(0.002)	\\
Constant	&	0.777***	&	1.299***	&	1.343***	&	1.337***	&	0.791***	&	0.931***	&	0.975***	&	0.976***	\\
	&	(0.017)	&	(0.019)	&	(0.020)	&	(0.019)	&	(0.021)	&	(0.009)	&	(0.009)	&	(0.009)	\\
Year dummies	&	yes	&	yes	&	yes	&	yes	&	no	&	no	&	no	&	no	\\
Observations	&	311,034	&	311,034	&	311,034	&	311,034	&	311,034	&	311,034	&	311,034	&	311,034	\\

\bottomrule
\multicolumn{9}{p{22.5cm}}{\textit{Notes:} The dependent variable is the growth rate of exports. The number of products is at 6-digit of the HS. Robust standard errors are in parentheses. Significance level: *** p$<$0.01, ** p$<$0.05, * p$<$0.10.}\\
\label{tab:estimations}
\end{tabular}
$  $}
\end{center}
\end{table}

In the different specifications estimated with OLS, the growth rate of the previous period has a negative effect in the growth of exports. Conversely, when considering the non-normality of the distribution of the data (models 5-8), the growth rates in $t-1$ are expected to have a positive effect on exports growth rates in $t$. This is the main difference between the results of the OLS and the ALAD estimation methods. However, despite the estimations of the autoregressive term report opposite signs using OLS and ALAD, the estimated coefficients are relatively small. 

In all the estimated models, smaller firms (that trade low quantities) are expected to grow more compared to larger firms (exports$_{(t-1)}$). This agrees with the fact that the right tail of the growth rates distribution allows for larger events for smaller firms (see Figure~\ref{fig:growth_binned}). 

Model (1) and (5) report the results for our benchmark model. We obtain robust estimations that indicate that diversification in both products and destinations have positive effects on the growth of exports in all the estimated models (1-8). Also, in all the estimated models, diversification in the destinations of exports increases trade growth rates more than diversification of products. 

The variables indicating the type of firm in terms of diversification and specialization in products and destinations can shed further light on these effects. Models (2) and (6) present the results using highly-diversified firms (multiple products and destinations) as the baseline. We observe that being a highly-specialized firm has a negative impact on growth rates compared to being a highly-diversified firm and also compared to firms that are specialized in products or destinations but diversified in either destinations or products. In general, all types of less diversified firms are associated with lower growth rates of exports compared to highly-diversified firms. Also, it is interesting to note that the difference is lower for the case in which firms are diversified in trade partners but specialized in products, which enriches the idea that diversifying in destinations generates higher growth opportunities than diversifying in products.

Models (2) and (6) also includes a dummy indicating whether the firm does processing trade. We estimate significant and positive coefficients for the dummy, which means that firms that do processing trade grow more than firms doing only ordinary trade.

In models (3) and (7), we also include variables indicating the ownership type of firms, using state-owned enterprises as the baseline. The results show that private-owned enterprises grow more than state-owned enterprises but foreign-owned enterprises grow less compared to both state-owned and private-owned enterprises. This is surprising to some extent but, foreign-owned enterprises are highly involved in processing trade (57.69\%), while only 27.05\% of state-owned enterprises and 9.80\% of private-owned enterprises do processing trade. Moreover, the shares of processing trade in total trade of each type of firms is quite different. While foreign-owned enterprises have a share of between 78 and 84\% of processing trade in their total exports, this share reaches only between 28 and 34\% for state-owned enterprises, and between 8 and 17\% for private-owned enterprises, depending on the year. Considering this, the dummy for processing trade might be capturing to a greater extent the behavior of foreign-owned enterprises. Thus, in models (4) and (8), we consider the interaction between ownership type and processing trade. The estimated coefficients show that private-owned and foreign-owned enterprises that do processing trade grow more than state-owned enterprises that do processing trade.

The lack of data characterizing firms prevents us from making a more thorough econometric analysis. However, the estimations using different specifications show that diversification in products and destinations of exports increase growth rates of exports. This is confirmed for different types of firms in terms of diversification in products and destinations: more diversified firms show higher growth rates while more specialized firms have lower grow rates. In particular, diversification in destinations has a stronger effect in exports growth rates. This implies that reaching a new destination leads to higher growth opportunities than starting a new production line. Moreover, facing a shock in the international market (for example, a global or regional crisis), it could be easier for a firm to redirect exports to a different destination than to reallocate resources to develop new or different products, especially, considering that industries usually present stickiness in the reallocation of resources from one sector to another.


\section{Concluding remarks}\label{conclu}

This paper studies trade patterns of Chinese firms for the period 2000 to 2006. The statistical analysis reveals that there exists high heterogeneity in Chinese exporting firms. The distribution of exports resembles a log-normal distribution. Many relatively small exporting firms coexist with a few very large exporting firms. Also, we observe fat tails in both sides of the distribution of growth rates, which implies that firms are often affected by extreme events. In addition, there is a negative relation between size and volatility of the growth rates. The distribution of exports growth rates shows that firms of different size (in terms of exports) are similarly affected by negative growth rates, but smaller firms have higher probabilities of growing at higher growth rates than medium and large firms. Finally, after trade liberalization, firms face more frequently stronger negative shocks.

The analysis of the changes in the intensive and extensive margins of exports showed that more than half of the firms intensified their exports between 2000 and 2006. But also, an important part of the firms diversified products (47.81\%) and destinations (54.76\%) enlarging their extensive margins. We showed that the exports of Chinese firms present economies of scope, which can explain, to a certain extent, the increase in the diversification of exported products and destinations of exports.

The econometric estimations showed a negative relation between size and growth, a positive impact of diversification in products and destinations on trade growth rates, a significant effect of different ownership types of firms, and that firms involved in processing trade also exhibit higher growth rates. The results are robust to different specification of the model and to different estimation methods.

Interestingly, highly-diversified firms grow more than highly-specialized firms and also than firms that diversified in products or destinations but specialized in destinations of products, respectively. 
Diversification in destinations have a stronger positive effect than diversification in products. This implies that facing more competition, firms might benefit more from exporting to a new destination rather than starting a new production line or reallocating resources to produce and export new products. In fact, we can consider that the development of a new exportable product can require more capabilities than reaching a new destination with an old product. Also, this suggests that, facing a negative shock, it could be easier for a firm to redirect exports to a different country than reallocating resources to develop new products.

\clearpage
\newpage
\bibliographystyle{chicago}
\bibliography{biblio}

\clearpage

\end{document}